\RequirePackage{lineno}
\documentclass[aps,prl,
twocolumn,showpacs,amssymb]{revtex4}
\usepackage{t1enc}
\usepackage[latin1]{inputenc}
\usepackage[english]{babel}
\usepackage{graphics}
\usepackage{graphicx}
\usepackage{epsfig}
\usepackage{amssymb}
\usepackage{dcolumn}
\usepackage{bm}
\usepackage{color}

\usepackage{lineno}
\begin{document}

\renewcommand{\figurename}{Fig.}

\title{Dissimilarities between the electronic structure of chemically doped and chemically pressurized iron pnictides from  an angle-resolved photoemission spectroscopy study}
\author{S.\ Thirupathaiah,$^1$  E.D.L.\ Rienks,$^1$  H.S.\ Jeevan,$^2$ R.\ Ovsyannikov,$^1$  E.\ Slooten,$^3$ J.\ Kaas,$^3$ E.\ van Heumen,$^3$ S.\ de Jong,$^{3,4}$ H.A.\ D\"urr,$^{1,4}$ K. Siemensmeyer,$^1$  R.\ Follath,$^1$   P.\ Gegenwart,$^2 $ M.S.\ Golden, $^3$ and J.\ Fink,$^{1,5}$}

\affiliation{
$^1$Helmholtz-Zentrum Berlin, Albert-Einstein-Strasse 15, 12489 Berlin, Germany\\
$^2$I. Physikalisches Institut, Georg-August-Universit\"at-G\"ottingen, 37077 G\"ottingen, Germany\\
$^3$Van der Waals-Zeeman Institute, University of Amsterdam, NL-1018XE Amsterdam, The Netherlands\\
$^4$Pulse Institute and Stanford Institute for Energy and Materials Science, SLAC National Accelerator Laboratory, Menlo Park, California 94025, USA\\
$^5$Leibniz-Institute for Solid State and Materials Research Dresden, P.O.Box 270116, D-01171 Dresden, Germany\\}
\date{\today}

\begin{abstract}
We have studied the electronic structure of EuFe$_2$As$_{2-x}$P$_x$ using high resolution angle-resolved photoemission spectroscopy. Upon substituting As with the isovalent P, which leads to a chemical pressure and to superconductivity, we observe a non-rigid-band like change of the electronic structure along the center of the Brillouin zone (BZ): an orbital and $k_z$ dependent
increase or decrease in the size of the hole pockets near the $\Gamma-Z$ line. On the other hand, the diameter of the Fermi surface cylinders at the BZ corner forming electron pockets, hardly changes.  This is in stark contrast to p and n-type doped iron pnictides where, on the basis of  ARPES experiments, a more rigid-band like behavior has been proposed. These findings indicate that there are different ways in which the nesting conditions can be reduced causing the destabilization of the antiferromagnetic order and the appearance of the superconducting dome.
\end{abstract}
\maketitle


The discovery of high T$_c$ superconductivity in ferropnictides \cite{Kamihara2008} has attracted a great deal of attention not only because of the high superconducting transition temperatures $T_c$ up to 55 K but also because of the complex phase diagrams. Starting from the antiferromagnetic (AF) metallic parent compounds, superconductivity  can be induced in various ways: by chemical doping, i.e., introducing electrons  or holes, by pressure, or by chemical pressure, e.g. in the AFe$_2$As$_2$ (A=Ca, Sr, Ba, and Eu) systems (122 systems) by substituting the As ions by isovalent but smaller P ions~\cite{Jiang2009}. All these methods lead to similar phase diagrams, i.e., the AF order is suppressed with increasing distance from the parent compound and a superconducting dome appears. The similarity of the phase diagrams hints at a common mechanism for this behavior. In a more itinerant picture, which is still under debate, these phase diagrams are explained in terms of decreasing nesting conditions between hole pockets in the center and electron pockets at the corner of the Brillouin zone (BZ)~\cite{Mazin2008a}. The similarity of the phase diagrams may indicate that in all cases the nesting conditions are reduced in the same way, i.e., the evolution of the electronic structure as a function of the distance to the parent compound is the same. On the other hand, the application of doping, pressure, or chemical pressure may be taking different routes to reach the same destination, inducing differing changes to the electronic structure.

In previous angle-resolved photoemission spectroscopy (ARPES) studies~\cite{Brouet2009,Thirupathaiah2010} a doping induced reduction of the nesting conditions have been reported: for n-type doped systems such as  BaFe$_{2-x}$Co$_x$As$_2$, a gradual decrease (increase) of the size of the hole (electron) pockets has been detected. For p-type doped systems such as Ba$_{1-x}$K$_x$Fe$_2$As$_2$, the opposite behavior has been observed. On the basis of these results, a doping induced shift of the Fermi level $E_F$, in  a rigid-band-like electronic structure has been proposed~\cite{Brouet2009,Thirupathaiah2010}. Theoretically, the ARPES results have been supported by density functional calculations (DFT) in the virtual crystal approximation and using super cells~\cite{Sefat2008,Thirupathaiah2010}. There is not, however, unanimity on the point of doping leading to a simple shift of the chemical potential. Recent super cell DFT calculations of different dopants (among other Co) on the Fe site indicated the formation of localized states leading to an isovalent substitution and not to a chemical doping~\cite{Wadati2010}.

Ba(Eu)Fe$_2$As$_{2-x}$P$_x$ \cite{Jiang2009,Ren2009a} is an ideal system to study the evolution of the electronic structure as a function of chemical pressure. In the Ba compound P substitution has led to superconductivity with a $T_c$ up to 30 K~\cite{Jiang2009}. In the Eu system, the AF order of the Fe ions is also suppressed by P substitution and the Eu ions order antiferromagnetically below 18 K. At higher P concentrations, however, the Eu moments order ferromagnetically which limits the superconducting dome to a narrow range between x=0.3 and 0.4~\cite{Jeevan2010}. To the best of our knowledge no ARPES investigations have been performed on these systems.  A series of recent articles on the electronic structure of AFe$_2$P$_2$ compounds \cite{Kasahara2010,Analytis2009,Coldea2009} using the de Haas-van Alphen (dHvA) effect have revealed that the three-dimensional (3D) nature of the electronic structure is increased compared to their As-based AF sister compounds. Furthermore, in a recent dHvA study on  BaFe$_2$As$_{2-x}$P$_x$ (0.41<$x$<1) a strong mass enhancement of the quasiparticles has been detected near the P concentrations that gives rise to the maximal $T_c$~\cite{Shishido2010}. Powerful as they are, these dHvA measurements do have the drawback that, due to the short mean free path of the hole carriers, it is difficult to identify and analyse the hole-like Fermi surfaces in the substituted compounds.

\begin{figure}[t]
	\centering
		\includegraphics[width=0.4\textwidth]{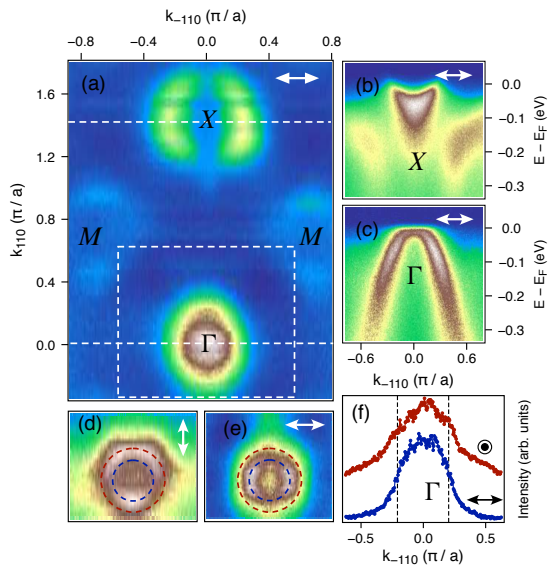}
		\caption{(color online) ARPES data of EuFe$_2$As$_{1.56}$P$_{0.44}$ taken with a photon energy $h\nu$ = 93 eV. (a) Fermi surface map along $\Gamma-X$ symmetry line measured with $s$-polarized light (indicated by a white double arrow). (b) and (c) energy distribution maps near $X$ and $\Gamma$, respectively. (d) and (e) constant energy contours 50 meV below $E_F$ in the $k$-range indicated in (a) by the rectangle marked with a white dashed line, for two different polarizations. (f) momentum distribution curves at $\Gamma$ near $E_F$ for two different polarizations.}
		\label{fig:Fig1}
\end{figure}

In this Letter we present a systematic study of the evolution of the electronic structure of the chemically pressurized system EuFe$_2$As$_{2-x}$P$_x$ using high-resolution ARPES to reveal the dispersion of bands parallel and perpendicular to the Fe layers. We observe non-rigid-band like shifts of bands related to the hole pockets. The changes in the electronic structure due to chemical pressure result in a charge neutral transfer of holes in the center of the BZ along the $k_z$ axis between bands having different orbital character. The size of the electron pockets at the corner of the BZ hardly changes. The changes in the center of the BZ signal an even stronger 3D electronic structure compared to the case of electron doping~\cite{Thirupathaiah2010}. Our data show that differing routes can and are taken to reach superconductivity in the ferropnictides: upon aliovalent substitution there are signs for a Fermi level shift in a more or less rigid band structure, whereas isovalent substitution--chemical pressure--gives a strongly non-rigid-band like evolution.  

Single crystals of EuFe$_2$As$_{2-x}$P$_x$ were grown in G\"ottingen using the Bridgman and Sn-flux method~\cite{Jeevan2008}. The AF transition temperature $T_N$ for the Fe$^{2+}$ moments in the parent compound was determined to be about 190 K while Eu$^{2+}$ moments order antiferromagnetically below T$_N$=18 K. A narrow superconducting dome with a maximum transition temperature $T_c$ = 29 K has been observed around x=0.4. Further sample characteristics will be published in a separate communication~\cite{Jeevan2010}. ARPES measurements were carried out at the BESSY II synchrotron radiation facility using the UE112-PGM2 beam line and ''$1^3$''-ARPES end station equipped with a Scienta R4000 analyzer. The total energy resolution was set between 5 meV to 7 meV while the angular resolution was 0.2$^\circ$ (0.3$^\circ$) parallel (perpendicular) to the analyser slit. High symmetry points of the BZ are denoted 
by $\Gamma$ = (0, 0, 0), Z = (0, 0, 1), 
X = (1/2, 1/2, 0), and K = (1/2, 1/2, 1) in units of (2$\pi/a$, 2$\pi/a$, 2$\pi/c$), where
$a$ and $c$ are the tetragonal lattice constants of EuFe$_2$As$_{2-x}$P$_x$. All measurements were performed below 20 K if not otherwise stated. Further experimental details have been published previously~\cite{Fink2009}.

Figure 1 shows representative ARPES data of EuFe$_2$As$_{1.56}$P$_{0.44}$, a slightly over-substituted compound. 
In the following we discuss the results and in particular the orbital character of the observed spectral weight in a coordinate system which is rotated by 45$^\circ$ around the $k_z$ axis. In this way it is easier to compare the data with calculations presented in Ref.~\onlinecite{Graser2009}. The map displayed in Fig.~1(a) shows an ellipsoidal Fermi surface (FS) formed by an electron pocket [see Fig.~1(b)] around the $X$ point, the intensity of which is strongly reduced along the $\Gamma-X$ line. This might be related to matrix element effects suppressing the intensity along a mirror plane for specific photon polarizations~\cite{Thirupathaiah2010}, although the intensity reduction along the $\Gamma-X$ line remains for p-polarized light, and for geometries free of mirror plane effects. According to Ref.~\onlinecite{Graser2009} the Fermi surface at $X$ should have $yz$ and $xy$ character perpendicular and parallel to the $\Gamma-X$ line, respectively. Both should be visible in our geometry for $s$-polarized light. Thus the $xy$ states would have to be strongly suppressed by matrix element effects other than the polarization selection rules. This view is supported by ARPES measurements on cuprates~\cite{Inosov2007} and by matrix element calculations for ferropnictides~\cite{Zhang2009}. Consequently, we state that the visible FS of the electron pocket near $X$ has $yz$ character. Next we discuss the spectral weight in Fig. 1(a) near $\Gamma$ which results from hole bands, the top of which are close to $E_F$ [see Fig.~1(c)]. Polarization dependent constant energy contours at 50 meV below $E_F$ shown in Figs. 1(d) and 1(e) indicate two almost degenerate hole pockets with an asymmetric intensity distribution around $\Gamma$ which can be nicely explained by the calculations of Ref.~\onlinecite{Graser2009} in terms of $xz/yz$ bands. The inner (outer) pocket shows more intensity for the $yz$ ($xz$) arcs perpendicular (parallel) to $\Gamma-X$ for $s$ ($p$) polarized light. In the momentum distribution curves shown in Fig.~1(f) we detect an outermost third hole pocket at $\Gamma$, highlighted by dashed lines, for p but not for s-polarized photons. From the low intensity and from the small $k_z$ dependence (see below) we tentatively assign this hole pocket to $xy$ states. The polarization dependence could be explained by a small matrix element for the odd $xy$ states and an admixture of even $x^2-y^2$ states. Summarizing the results derived from Fig. 1, our results on the orbital character of the states close to $E_F$ at $\Gamma$ and $X$ are compatible with the theoretical results presented in Ref.~\onlinecite{Graser2009}.

To elucidate the 3D nature of the electronic structure we have measured $h\nu$ dependent ARPES data at the center and at the corner of the BZ (see Fig. 2). An inner potential of 15 eV has been used to calculate the $k_z$ values from $h\nu$ with the $c$ values reported in Ref.~\onlinecite{Ren2009a}. The $\Gamma, Z, X,$ and $K$ points are indicated by dashed lines. Along the $\Gamma-Z$ direction, we compare data taken with two different polarizations, so as to enable the extraction of important information regarding the orbital character of the states near $E_F$. For $s$-polarized light, in this geometry, for $k_{-110}=0$ we can detect only odd states relative to the $\Gamma-X$ mirror plane and therefore the $k_z$ dispersive band in Fig. 2(a) has $yz$ character. For $p$-polarized photons, we are sensitive to even states, which have $xz$, $x^2-y^2$, and $z^2$ character. Thus we identify the inner slightly dispersing Fermi cylinders with the almost degenerate $xz/yz$ bands, different from our previous assignment~\cite{Thirupathaiah2010}. The almost non-dispersive next in diameter size cylinder is assigned to $xy$ states, visible by the admixture of even $x^2-y^2$ states. In agreement with our previous results~\cite{Thirupathaiah2010} we assign the outermost Fermi cylinder, which appears near the $Z$ point, to $z^2$ states. As is to be expected, the $k_z$ dispersion is large for orbitals not laying in the Fe plane and small for in-plane orbitals. The data near $k_x,k_y=0$ indicate a considerably large $k_z$ dispersion, i.e., a larger three-dimensionality when compared to the non-substituted compound. At the zone corner we observe a small $k_z$ dispersion of the Fermi cylinder.
\begin{figure}[t]
	\centering
		\includegraphics[width=0.4\textwidth]{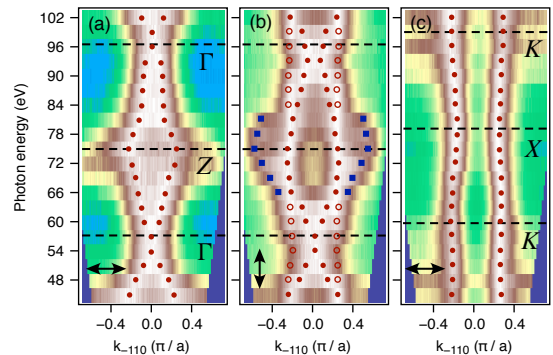}
		\caption{(color online) ARPES Fermi surface maps of EuFe$_2$As$_{1.56}$P$_{0.44}$ in the $k_{001}-k_{-110}$ plane. Actually the photon energy dependence of the intensity is shown. (a) and (b) data near the $\Gamma-Z$ line taken with $s$ and $p$ polarized light, respectively. (c) data near the corner of the BZ measured with $p$-polarized light. The symbols indicate the band dispersion having different orbital character: open (closed) circles predominantly $xy$ ($xz/yz$) character, closed squares predominantly $z^2$ character.}
	\label{fig:Fig2}
\end{figure}

Analogous measurements as those shown in Figs. 1 and 2 have been performed for the P concentrations $x$= 0.28 and 0.32. The size of the hole pocket in the parent compound was derived from data taken above $T_N$. From the entirety of these data we derive the Fermi vectors, $k_F$, along the $k_{-110}$ direction for the high symmetry points $\Gamma, Z, X$, and $K$ as a function of P substitution, which are presented in Fig.~3(a). We observe a non-rigid-band evolution of the electronic structure upon P substitution. At $\Gamma$ the size of the $xz/yz$ hole pockets  decrease with $x$, whereas that of the $xy$ pocket remains constant. At $Z$ the size of the $z^2$ hole pocket increases and that of the almost degenerate $xy, xz/yz$ hole pockets slightly increase. The size of the electron pockets at $X$ and $K$ hardly changes. For $k_F\gg dk_F$ and for a 2D electronic structure there is a linear relationship between the concentration dependent changes of the total number of charge carriers $dn_{e,h}/dx$ and the concentration dependence of the Fermi vector $dk_F/dx$:
\begin{equation}
dn_{e,h}/dx= 2\pi k_F(dk_F/dx)/S_{BZ},
\end{equation}
where $S_{BZ}=(2\pi/a)^2$ is the area of the BZ. Using the data of Fig. 3, for $x\leq$ 0.32 we derive that the increase of $k_F(dk_F/dx)$ for the $z^2$ holes at Z is about the same as the decrease of the same value for the two $xz/yz$ hole pockets at $\Gamma$. Remembering that the other pockets do not change this signals a charge neutral replacement of the As atoms by P, as is to be expected for isovalent substitution. For $x$=0.44 the data could indicate an apparent deviation from charge neutrality which needs further investigations.

The non-rigid-band evolution of electronic structure is probably related to a change of the crystal field splitting of the Fe 3$d$ states upon substitution of the larger As ions by  the smaller isovalent P ions thus changing the pnictogen height above the Fe layers. The $k_F$ data for EuFe$_2$As$_{2-x}$P$_x$  indicate a decreasing nesting condition with increasing P substitution since part of the hole pockets at $\Gamma$ and $Z$ decreases and increases, while the size of the electron pocket is almost constant.
\begin{figure}[t]
	\centering
		\includegraphics[width=0.4\textwidth]{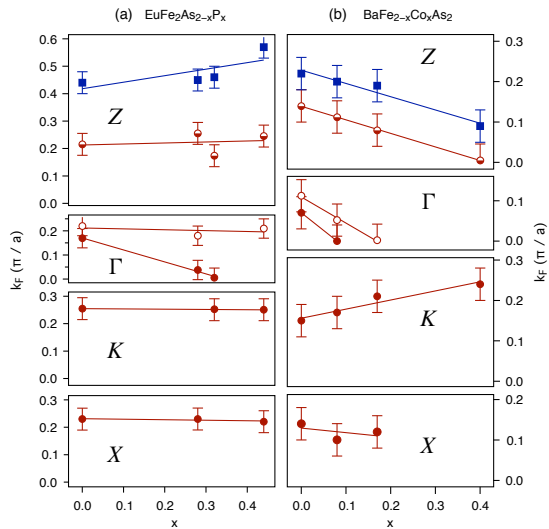}
		\caption{(color online) Fermi vectors plotted as a function of doping/substitution concentrations at $Z$, $\Gamma$, $K$, and $X$. (a) isovalently substituted (chemically pressurized) system EuFe$_2$As$_{2-x}$P$_x$. (b) Aliovalently substituted (electron doped) BaFe$_{2-x}$Co$_x$As$_2$.  The assignment of the symbols to the orbital character is the same as in Fig. 2.}
	\label{fig:Fig3}
\end{figure}

We contrast the $k_F$ values for EuFe$_2$As$_{2-x}$P$_x$ with analogous values [see Fig.~1(b)] for the chemically doped system BaFe$_{2-x}$Co$_x$As$_2$ which were taken from the data presented in Ref.~\onlinecite{Thirupathaiah2010}. In this case both the hole pockets at $\Gamma$ and at $Z$ decrease with increasing Co concentration while the electron pocket increases. These results also indicate a decrease of the nesting conditions with increasing Co concentration. The observed $k_F$ values as a function of Co concentration could be explained by a doping induced shift of the Fermi level to higher energies upon n-type doping. In the following, we approximate the 3D electronic structure by a 2D one, in which the size of Fermi surfaces is calculated from the average of the $k_F$ values at $k_z$ = 0 and 1, and we assume circular Fermi surfaces. Using Eq. (1) and the $k_F$ and $dk_F/dx$ values from Fig. 3(b) we derive an increase of 0.7 electrons per Co atom.

It is interesting to note that in both systems, the Co and the P substituted compounds, for $k_z$=0 and near optimal doping (highest $T_c$) we observe a closing of the $xz/yz$ hole pockets. This would lead to a reduction of the nesting conditions between hole and electron pockets and possibly makes the nesting between the electron pockets more important. In this way it would be possible to understand doping dependent changes of the symmetry of the superconducting order parameter~\cite{Tanatar2010}, e.g. from s$\pm$ to d. A similar scenario was described in Ref.~\onlinecite{Kuroki2009}. On the other hand a rapid decrease of the nesting conditions above a certain  concentration is rather unlikely when looking at the data for $k_z\neq 0$. For these $k_z$ values the $xz/yz$ hole pockets exist over a much larger concentration range which means they are available for nesting even well above the optimal doping/substitution concentration. 

In the following we compare the present results with data previously published in the literature. During our manuscript preparation we noticed a similar dHvA study on BaFe$_2$(As$_{0.37}$P$_{0.63}$)$_2$ in which an enhanced nesting for the superconducting compound in comparison to the fully substituted compound was derived~\cite{Analytis2010}. We
point out that in these measurements only one instead of three hole pockets have been detected and therefore on the basis of these data it is difficult to conclude  on the nesting conditions. For the system BaFe$_{2-x}$Co$_x$As$_2$ various ARPES studies have been published~\cite{Brouet2009,Thirupathaiah2010}. In general these ARPES results indicate a rigid-band like behavior upon n-type doping with Co (and for p-type doping by replacement of Ba by K). Our value 0.7 electrons transfered from Co to the Fe-derived bands is close to the value of 1 from Ref.~\onlinecite{Brouet2009}. From the fact that we have not measured the full Fermi surfaces for all $k_x$, $k_y$, and $k_z$ values we cannot rule out a complete
charge transfer from Co to the Fe-derived bands.

In summary we have performed high-resolution  ARPES studies on EuFe$_2$As$_{2-x}$P$_x$ in order to reveal the nature of the electronic structure as a function of $x$. The results are compared with analogous data of the electron doped BaFe$_{2-x}$ Co$_x$As$_2$ system. We conclude that the evolution of electronic structure upon substitution of As by P in EuFe$_2$As$_2$, which leads to chemical pressure, is non-rigid-band like in nature. On the other hand, for the aliovalently substituted system BaFe$_{2-x}$Co$_x$As$_2$ we see more signs compatible with a shift of $E_F$ in a rigid-band like electronic structure. Our findings are supporting the importance of nesting conditions for the understanding of the phase diagram, the appearance of superconductivity and the pairing symmetry of the order parameter in ferropnictide compounds.

Financial support by the DFG through SPP1458 and from FOM(NWO)is gratefully acknowledged.

\bibliographystyle{phaip}
\bibliography{BaFe2As2-xPx}

\begin{thebibliography}{10}

\bibitem{Kamihara2008}
Y.~Kamihara et~al.,
\newblock J. Am. Chem. Soc. {\bf 130}, 3296 (2008).

\bibitem{Jiang2009}
S.~Jiang et~al.,
\newblock J. Phys.: Condens. Matter {\bf 21}, 382203 (2009).

\bibitem{Mazin2008a}
I.~I. Mazin et~al.,
\newblock Phys. Rev. Lett. {\bf 101}, 057003 (2008).

\bibitem{Brouet2009}
V.~Brouet et~al.,
\newblock Phys. Rev. B {\bf 80}, 165115 (2009).

\bibitem{Thirupathaiah2010}
S.~Thirupathaiah et~al.,
\newblock Phys. Rev. B {\bf 81}, 104512 (2010).

\bibitem{Sefat2008}
A.~S. Sefat et~al.,
\newblock Phys. Rev. Lett. {\bf 101} (2008).

\bibitem{Wadati2010}
H.~Wadati et~al.,
\newblock arXiv:1003.2663  (2010).

\bibitem{Ren2009a}
Z.~Ren et~al.,
\newblock Phys. Rev. Lett. {\bf 102}, 137002 (2009).

\bibitem{Jeevan2010}
H.~Jeevan and P.~Gegenwart,
\newblock (unpublished) .

\bibitem{Kasahara2010}
S.~Kasahara et~al.,
\newblock Phys. Rev. B {\bf 81}, 184519 (2010).

\bibitem{Analytis2009}
J.~G. Analytis et~al.,
\newblock Phys. Rev. Lett. {\bf 103}, 076401 (2009).

\bibitem{Coldea2009}
A.~I. Coldea et~al.,
\newblock Phys. Rev. Lett. {\bf 103}, 026404 (2009).

\bibitem{Shishido2010}
H.~Shishido et~al.,
\newblock Phys. Rev. Lett. {\bf 104}, 057008 (2010).

\bibitem{Jeevan2008}
H.~S. Jeevan et~al.,
\newblock Phys. Rev. B {\bf 78} (2008).

\bibitem{Fink2009}
J.~Fink et~al.,
\newblock Phys. Rev. B {\bf 79}, 155118 (2009).

\bibitem{Graser2009}
S.~Graser et~al.,
\newblock New J. Phys. {\bf 11}, 025016 (2009).

\bibitem{Inosov2007}
D.~S. Inosov et~al.,
\newblock Phys. Rev. Lett. {\bf 99}, 237002 (2007).

\bibitem{Zhang2009}
Y.~Zhang et~al.,
\newblock arXiv:0904.4022  (2009).

\bibitem{Tanatar2010}
M.~A. Tanatar et~al.,
\newblock Phys. Rev. Lett. {\bf 104}, 067002 (2010).

\bibitem{Kuroki2009}
K.~Kuroki et~al.,
\newblock Phys. Rev. B {\bf 79}, 224511 (2009).

\bibitem{Analytis2010}
J.~Analytis et~al.,
\newblock arXiv:1002.1304  (2010).

\end{thebibliography}
\end{document}